\documentclass[12pt,a4paper]{article}
\usepackage[pdftex]{graphicx}
\usepackage{jheppub}
\usepackage{hyperref}
\usepackage{amsfonts, amsmath, amssymb}

\usepackage{esint} 
\usepackage{geometry} 
\usepackage{bbm} 
\usepackage{slashed} 

\DeclareRobustCommand\openone{\leavevmode\hbox{\small1\normalsize\kern-.33em1}}%

\newcommand{\nn}{~,\nonumber\\}
\newcommand{\ket}[1]{\vert #1 \rangle}
\newcommand{\bra}[1]{\langle #1 \vert }
\newcommand{\beas}{\begin{eqnarray*}}
\newcommand{\eeas}{\end{eqnarray*}}
\newcommand{\bes}{\begin{equation*}}
\newcommand{\ees}{\end{equation*}}
\newcommand{\be}{\begin{equation}}
\newcommand{\ee}{\end{equation}}
\newcommand{\bea}{\begin{eqnarray}}
\newcommand{\eea}{\end{eqnarray}}
\newcommand{\Res}{\mathop{\mathfrak{Res}}}
\newcommand{\e}[1]{ \textrm{\large e}^{\displaystyle #1}  } 

\newtheorem{proposition}{Proposition}

\begin{document}
\rightline{\vbox{\small\hbox{\tt NITS-PHY-2011007} }}
\vskip 1.8 cm

\title{Quantum electrodynamics of spin $3/2$, II}
\author{Konstantin G. Savvidy}
\affiliation{Physics Department, Nanjing University, Hankou Lu 22, Nanjing, P.R. of China}
\abstract{
Electromagnetic interactions of the spin 3/2 particle are investigated while allowing the propagation of the transverse spin 1/2 component present  in the reducible Rarita-Schwinger vector-spinor. This is done by allowing a more general form for the mass term, while leaving the kinetic terms untouched. We find that the interaction is consistent and does not lead to superluminal propagation for a range of the mass of the spin 1/2 particle, except for the special point where the spin 1/2 particle is infinitely massive. 
We then quantize the theory using the appropriate Grassmann-variable path integral and study the poles of the propagator. The unfamiliar feature of the theory is that the charge matrix is not positive definite; it is positive definite on the space of spin 3/2 solutions, and negative definite on the space of spin 1/2 solutions. Classically, for spin 1/2 modes the energy is of the opposite sign to frequency, i.e. the positive frequency modes have negative energy. It is found that the retarded part appropriately propagates the positive energy solutions forward in time, while the advanced part propagates the negative energy solutions back in time so long as the parameters are chosen such that the positive frequency modes have the same parity. The propagator contains poles of the spin 1/2 modes above the real axis for the positive and below the real axis for the negative frequency modes, while the residues at those poles are also of the sign opposite to the usual; altogether this leads to a unitary S-matrix, the forward amplitude $\langle \mathbf{p} | \mathbf{p} \rangle $ being positive for all modes.  The canonically quantized field is  causal, and the equal-time anti-commutator has positive definite form.
In the canonical formalism it is only necessary to choose the physical vacuum as the highest-weight state with respect to the spin 1/2 operators in order to correctly reproduce both the unusual residues and locations of the poles.
%
}

\emailAdd{ksavvidis@gmail.com}

\maketitle
\flushbottom
\tableofcontents

\setlength{\arraycolsep}{2pt}

Recently, an interacting theory of higher-spin gauge fields, satisfactory at least at the classical level has been constructed by G. Savvidy \cite{Savvidy:2005fi,Savvidy:2005zm,Savvidy:2005ki}. The theory is free of ghosts \cite{Savvidy:2009zz}, and does not contain higher derivative non-renormalizable terms. The interactions are entirely of power-counting renormalizable cubic and quartic vertexes of the Yang-Mills type, and are dictated by a peculiar new extended gauge invariance principle. The new gauge invariance principle indeed unifies the space-time Lorentz and local gauge symmetry, formally going around the assumptions of Coleman and Mandula by containing an infinite number of fields. 

Also, it has been demonstrated that open string theory contains higher spin gauge bosons  with interactions of the Yang-Mills type \cite{Savvidy:2008ks}.  Presence of higher spin matter and gauge particles is a generic feature of string theory and sharply contrasts with the generic prediction of Kaluza-Klein-type theories with partners of the same spin and supersymmetry with partners of spin lower by 1/2. Quantum numbers of any new particles are highly constrained, for example by the requirement of anomaly cancellation, however it may be very difficult to measure the spin of such partner particles by direct measurement \cite{Cheng:2002ab}. 
Therefore, it is important to develop quantitative theories which can distinguish between supersymmetric, Kaluza-Klein and higher-spin partners.

In light of these developments, it becomes urgent to develop phenomenologically realistic extensions of the Standard Model, incorporating the new gauge theory of higher spin. On this path, the major obstacle is the obscure status of interacting half-integer spin fermionic matter fields. On the one hand, satisfactory free theories of higher-spin fermions are known \cite{fierz,fierzpauli,Singh:1974qz,fronsdal}. On the other hand, the extended gauge invariance principle allows to fix most of the ambiguity in the interactions of matter with gauge fields. In the present work  the details of interactions of spin 3/2 matter field with spin 1 abelian gauge field are considered. This problem is also interesting in its own right, without reference to the new theories of higher-spin gauge fields.

The equations analyzed in this paper are almost exactly those of the standard Rarita-Schwinger spin 3/2 theory \cite{Rarita:1941mf}. The difference is that we relax one of the Pauli-Fierz constraints, which are necessary to get rid of the reducible spin 1/2 components. The Rarita-Schwinger vector-spinor contains two spin 1/2 irreducible components, the undesirable longitudinal $ p_\mu \, \psi^\mu$ and the physical transverse $\gamma_\mu \, \psi^\mu$. 
We then find in Sec. \ref{sec:modes} that the theory contains one spin 3/2 and one spin 1/2 positive-norm physical particle, together with their antiparticles, while the other, longitudinal spin 1/2 component does not propagate, i.e. does not go on-shell for any value of the momentum. The kinetic terms are of the Rarita-Schwinger type, but adjusted to bring the theory into the form where the propagating fields are completely transverse, $ p_\mu \, \psi^\mu =0 $ for all propagating modes. The mass term of this theory,  $m \, (\delta^\mu_{~\nu} - z \, \gamma^\mu \, \gamma_\nu)$ contains the only free parameter, which we call $z$, and which defines the ratio of the masses of the two physical particles as $M_{1/2}/M_{3/2} =  \tfrac {1}{2\,(3\,z-1)}$. 

The theory is then investigated as a function of this free parameter. The value of the parameter that corresponds to the standard choice $z=1/3$, whereby the spin 1/2 particle becomes infinitely massive is untenable due to the appearance of superluminal propagation, as found by Velo and Zwanziger \cite{Velo:1969bt,Velo:1970ur}. We confirm in Sec. \ref{sec:super} that in the allowed range of the parameter superluminal propagation does not occur. For most of the allowed range of the parameter, the spin 1/2 particle is lighter than the spin 3/2 particle, and this is also the phenomenologically preferred arrangement. The solution to the Velo-Zwanziger problem has been suggested in substantially the same form as in the present work by Ranada and Sierra \cite{Ranada:1980yx}.  

Path integral quantization of the theory is considered in Sec. \ref{sec:prop}.
The propagator of the theory is found by inverting the kinetic operator in momentum space. In order to write it in a usable form it is necessary to build up the spin-sums to the separate physical subspaces; this is achieved by means of the well-known projection operators which separately project onto the transverse spin 3/2 and spin 1/2 spaces. 
Further, we make use of the appropriate parity operator in order to correctly project to the positive and negative frequency solutions. It is found that the retarded part of the propagator appropriately propagates the solutions with classically positive energy forward in time, while the advanced part propagates the negative-energy solutions back in time, so long as the parameter is chosen such that all positive frequency modes have the same parity. The propagator contains poles of the spin 1/2 modes above the real axis for the positive and below the real axis for the negative frequency modes, while the residues at those poles are also of the sign opposite to the usual; altogether this leads to a unitary S-matrix, the forward amplitude $\langle \mathbf{p} | \mathbf{p} \rangle $ being positive for all modes.
The propagator also satisfies the Ward identity of the form appropriate for this theory. 

In Sec.  \ref{sec:field} we consider the canonical quantization of the theory and present the second quantized field. 
The fully covariant field anti-commutator is then computed,  and is shown to vanish at space-like separations. At equal-times the right hand side of the anti-commutator is positive definite and consists of a sum of matrixes of positive unit eigenvalues. 
The vacuum is chosen  such as to make the energy of physical excitations positive, which is always possible in a fermionic theory. The unusual feature is that due to the non-definiteness of the charge matrix, it is necessary to choose the highest-weight vacuum in the spin 1/2 sector. Further, the two-point function is computed as the time-ordered product, and the profound consequences of choosing the highest-weight vacuum become apparent here: in the spin 1/2 sector, the negative frequency modes contribute to the retarded part, while the positive frequencies contribute to the advanced part.
Finally, we discuss what if any principle may allow the extrapolation from the on-shell spin sums to the off-shell propagator in the canonical formalism, since this is necessary to fix the form of the non-pole terms. With this caveat, the result  
is compared with the momentum space propagator which we already obtained in using the path integral in Sec \ref{sec:prop} and agreement is found. 

Since the arrangement of the poles and residues is so unusual, we present a general treatment of the Kallen-Lehmann spectral representation for fermions in Appendix \ref{sec:app2}. It is argued that unlike the bosonic case, where the highest-weight vacuum $ a^\dagger \, \ket{\Omega}=0$ leads to negative norm states, the highest-weight  vacuum in the fermionic theories is allowed. Once this is established, it inevitably follows that the negative frequency modes contribute to the retarded part of the propagator and the structure of poles and residues is naturally reproduced in the general setting.

The theory is of interest also for phenomenological reasons.
Calculations of Compton scattering with charged spin 3/2 particles in the intermediate state  arise in the phenomenological theory of photon-nucleon scattering \cite{Peccei:1969sb,Benmerrouche:1989uc,Pascalutsa:1995vx}, whenever the energy of the impinging gamma quantum is in the region of the $\Delta^+$ resonance, $\approx 200-300 MeV$. 
This remains an active subject of research \cite{Pascalutsa:1999zz, Pascalutsa:2006up, Lensky:2009uv}, and
our theory has the potential to bring clarity to some of the long-standing paradoxes and inconsistencies encountered in the investigation of this phenomenon. Specifically, we propose to describe the proton and the $\Delta^+$ as two different components of a single reducible Rarita-Schwinger multiplet, both particles really being the different states of the same physical system composed of the same three quarks. The present theory has physically acceptable poles in the propagator only if the physical particles of the same charge have the same intrinsic parity, which is known to be the case for the proton and the $\Delta^+$. Physical attractiveness of the unified description of the proton and $\Delta^+$ is  particularly convincing evidence in favor of allowing a spin 1/2 particle to propagate as part of the reducible Rarita-Schwinger vector-spinor representation. An application of the present theory to proton Compton scattering requires a study of the non-minimal couplings in the interacting theory. This work is well in progress, and will be published separately. 

\section{The Rarita-Schwinger Lagrangian.}
\label{sec:rs}
There exists adequate formulation in the literature for the free equations of massive and massless fermions \cite{fierz,fierzpauli, Singh:1974qz, fronsdal}. Subsequently, it was discovered by Velo and Zwanziger \cite{Velo:1969bt,Velo:1970ur} that whenever the massive particle is minimally coupled to the electromagnetic field, there exist solutions with superluminal propagation. Nevertheless, some calculations are unaffected by such difficulties in the formal theory, for example in the same
issue of Physics Review as the Rarita and Schwinger article there appeared an analysis by Kusaka \cite{Kusaka:1941} ruling out the possibility that the neutrino is a spin 3/2 particle described by the newly found equations. 

The form of the spin $3/2$ lagrangian we would like to study is
\be
\label{eq:lagr}
L =  - \bar{\psi}_\mu ~ [D^\mu_{~\nu} - m\, \Theta^\mu_{~\nu}]~ \psi^\nu~, 
\ee
where the kinetic term $D$ is strictly linear in momentum and $\Theta$ is a mass term independent of momentum. Both are Lorentz tensors of the second rank:
\begin{eqnarray}
\label{eq:kin}
D^\mu_{~\nu} &=& (\gamma^\rho \, p_\rho)\, \delta^\mu_{~\nu}  
+ \xi \, (\gamma^\mu \, p_\nu +  \gamma_\nu \, p_\mu)
+ \zeta \, (\gamma^\mu \, \gamma^\rho \, p_\rho \, \gamma_\nu) \nn
 \Theta^\mu_{~\nu} &=& \delta^\mu_{~\nu} - z \, \gamma^\mu \, \gamma_\nu~\nn
\zeta&=&(3\,\xi^2+2\,\xi+1)/2\nn
\xi &=& 2 \, z -1~.
\end{eqnarray}
All terms allowed by criteria mentioned are present in these tensors, however relative coefficients in $D^\mu_{~\nu}$ have been already tuned to the Rarita-Schwinger-Chang-Hagen-Fronsdal form.
The kinetic term is more or less unique up to contact transformations, a well flagged point in the existing literature. We have in fact used a contact transformation in order to make the solutions of the equations of motion completely transverse, $ p_\mu \, \psi^\mu = 0$, by tuning the parameter $\xi$ depending on $z$. Therefore   $z$ is the only remaining free parameter.

The arbitrariness of $z$ had been previously dealt in the literature by the following considerations. The vector-spinor field $\psi^\mu$ is a reducible representation of the Lorentz group. Having 16 components, it contains $4+4=8$ degrees of freedom necessary to describe the spin 3/2 particle and its antiparticle. The remaining components fall into two Dirac spinors with 4 components each. 
Pauli and Fierz project out those, by imposing the conditions  $ p_\mu \, \psi^\mu = 0$ and $\gamma_\mu \, \psi^\mu=0$ by hand, while Rarita and Schwinger achieve this as a consequence of the equations of motion following from the lagrangian above with $z=1/3$.


\section{Propagating modes}
\label{sec:modes}
We proceed to investigate the solutions of the equations of motion $[D^\mu_{~\nu} - m\, \Theta^\mu_{~\nu}]~ \psi^\nu = 0$, while keeping $z$ as a free parameter.

The solutions with spin 3/2 do not depend on the parameter $z$, and are well-known. With gamma matrices in the Weyl representation, four-vector polarization vectors 
$E_1= (0,1, i,0)$,  
$E_2= (0,1,-i,0)$,
$E_3= (0,0,0,1)$,
$E_4= (1,0,0,0)$, and a basis of Dirac spinors 
$$
U_1={\begin{pmatrix}1\\ 0\\ 1\\ 0 \end{pmatrix} }~, 
~~ U_2={ \begin{pmatrix}0\\ 1\\ 0\\ 1  \end{pmatrix} }~,~~
V_1=\gamma_5\,U_1 = { \begin{pmatrix}1\\ 0\\ {-}1\\ 0  \end{pmatrix} }~,
 ~~ V_2=\gamma_5\,U_2 ={ \begin{pmatrix}0\\ 1\\ 0\\ {-}1  \end{pmatrix} }~.
$$
we may build up the positive frequency solutions with $\omega=m$ in the rest frame as
\begin{eqnarray}
\label{eq:e4}
u^\mu_4(0, \, +3/2)  &=& \tfrac{1}{2} \, E^\mu_1 \otimes U_1\nn
u^\mu_4(0, \, +1/2)  &=& \tfrac{1}{2\sqrt{3}} \,  E^\mu_1 \otimes U_2 - \tfrac{1}{\sqrt{3}} \,   E^\mu_3 \otimes U_1\nn
u^\mu_4(0, \, -1/2)  &=&  \tfrac{1}{2\sqrt{3}} \,  E^\mu_2 \otimes U_1 + \tfrac{1}{\sqrt{3}} \,   E^\mu_3 \otimes U_2\nn
u^\mu_4(0, \, -3/2)  &=& \tfrac{1}{2} \, E^\mu_2 \otimes U_2
\end{eqnarray}
The solutions are orthogonalized, with respect to each other, and in addition made to diagonalize the Rarita-Schwinger  spin operator
$\Sigma_3 =  \tau^{12} \otimes  \openone + \openone \otimes \sigma^{12} $, 
where $\tau$ and $\sigma$ are generators of rotations for ordinary vectors and spinors respectively. Thus, the solutions fill out the spin multiplet appropriate for a fermion of spin 3/2, and mass $M_{3/2}=m$.

Corresponding solutions with negative frequency $\omega=-m$ can be obtained by acting with $\gamma_5$, or equivalently, replacing $U$ for $V$ in \eqref{eq:e4}:
\begin{eqnarray}
\label{eq:p4}
v^\mu_4(0,+3/2)  &=& \tfrac{1}{2} \, E^\mu_1 \otimes V_1\nn
v^\mu_4(0,+1/2)  &=& \tfrac{1}{2\sqrt{3}} \,  E^\mu_1 \otimes V_2 -  \tfrac{1}{\sqrt{3}} \,  E^\mu_3 \otimes V_1\nn
v^\mu_4(0,-1/2)  &=&  \tfrac{1}{2\sqrt{3}} \,  E^\mu_2 \otimes V_1 +  \tfrac{1}{\sqrt{3}} \,  E^\mu_3 \otimes V_2\nn
v^\mu_4(0,-3/2)  &=& \tfrac{1}{2} \, E^\mu_2 \otimes V_2.
\end{eqnarray}
These are appropriate for the spin 3/2 anti-fermion. In the general frame these should be defined in the usual convention with 
positive $p_0$, so that the solution is $v(\mathbf{p}, s) \, e^{ipx}$ in contrast to the positive frequency modes which are $u(\mathbf{p}, s) \, e^{-ipx}$.

The operator $D^\mu_{~\nu} - m \, \Theta^\mu_{~\nu} $ is acting on the 16-component vector-spinors and as such is generally expected to have 16 eigenvalues and eigenvectors. Instead, we are interested in homogeneous solutions to the equations, and thus need to investigate the \emph{nullspace} of the operator for different values of momenta. Because we are interested in massive solutions, it is convenient to consider particles in the rest frame, $p=(\omega, 0,0,0)$. The characteristic polynomial is of the \emph{twelfth} degree in frequency $\omega$:
\newcommand{\Det}{\mathop{\mathrm{Det}}}
$$\Det |D^\mu_{~\nu} - m \, \Theta^\mu_{~\nu}| = [m + \omega]^4 \, [m - \omega]^4 \, 
 [M+\omega]^2 \, [M - \omega]^2~ m^4\, \left(\frac {4 \, z-1}{3\, z -1} \right)^4    ~,
$$
where $M=  \frac {m}{2\,(3\, z - 1)}$.
Thus, $4+4=8$ eigenvalues can be nullified by setting $\omega=\pm m$ as above. Another $4$ eigenvalues can be made equal to zero at $\omega=\pm M$, two for each sign. In principle, there is as yet no restriction on the sign of $M$, for negative values of $M$ the positive and negative frequency   solutions are simply interchanged.
However, as we shall see in Sections \ref{sec:prop} and \ref{sec:field} the additional restriction $M>0$ must be imposed in order to preserve unitarity, this restricts the range of $z$ to $z \in (1/3, \infty )$. In the usual Dirac theory there is no penalty for choosing a negative mass, and if we decided to choose $m<0$, then we must also have $M<0$ (with the same allowed range for $z$). In all cases it must be the case that the solutions with same frequency have the same parity. 

The corresponding solutions are appropriate for describing a physical particle of spin $1/2$ and mass 
$M_{1/2} =  \vert M \vert = \vert \frac {m}{2\,(3\, z - 1)}\vert$:
\begin{eqnarray} 
\label{eq:e2}
u^\mu_2(0,+1/2)  &=& {1/\sqrt{3}} \,  \left(E^\mu_1 \otimes U_2 + E^\mu_3 \otimes U_1 \right)\nn
u^\mu_2(0,-1/2)   &=& {1/\sqrt{3}} \,  \left(E^\mu_2 \otimes U_1 - E^\mu_3 \otimes U_2  \right)
\end{eqnarray}
and antiparticle
\begin{eqnarray} 
\label{eq:p2}
v^\mu_2(0,+1/2)  &=& {1/\sqrt{3}} \,  \left(E^\mu_1 \otimes V_2 + E^\mu_3 \otimes V_1  \right)\nn
v^\mu_2(0,-1/2)   &=& {1/\sqrt{3}} \,  \left(E^\mu_2 \otimes V_1 - E^\mu_3 \otimes V_2  \right)
\end{eqnarray}
These solutions are completely transverse, $p_\mu\, \psi^\mu =0 $. On the basis of this, one may expect that the corresponding quantum theory may be unitary.

The standard choice $z=1/3$, which corresponds to the original model of Rarita and Schwinger with $\xi=-1/3$, $\zeta= 1/3$ leads to the disappearance of the spin 1/2 states from the theory simply because the mass of those states goes to infinity  $M= \frac {m}{2\,(3\, z - 1)} {\rightarrow  } \infty $ as ${z\rightarrow1/3}$. 

The remaining four-dimensional vector subspace 
does not go on-shell for any value of frequency, because the characteristic polynomial is of the twelfth degree and allows only 12 propagating modes in \eqref{eq:e4}, \eqref{eq:p4} and \eqref{eq:e2},\eqref{eq:p2}. This is of course the result of the fine tuning inherent in the Rarita-Schwinger-Chang-Hagen-Fronsdal kinetic operator $D^\mu_{~\nu} $. We already demonstrated that there is a form of the theory where the solutions occupy only the twelve spatial components $\psi_{i\alpha}$. In fact, the lagrangian can be transformed by contact transformation to a form whereby it does not contain time derivatives of $\psi_0$ at all and this component then plays the role of a lagrange multiplier. 


This completes the construction of the 12 propagating modes corresponding to a spin 3/2 ($4+4=8$ modes) and a spin 1/2 ($2+2=4$ modes) physical particle  in the rest frame, and thus by boosting in any frame.

\section{Electrodynamics}
\label{sec:ed}
The interaction with an abelian gauge field can be introduced in the usual way by covariant derivative $p_\mu \rightarrow \pi_\mu=p_\mu -  A_\mu$. This ensures gauge invariance and conservation  of the current:
\be
\label{eq:current}
- j^\mu = \frac{\delta L}{\delta A_\mu} = 
\bar{\psi}_\nu \, \gamma^\mu \, \psi^\nu 
+ \xi \, (\bar{\psi}_\nu \, \gamma^\nu \psi^\mu + \bar{\psi}^\mu \gamma_\nu \, \psi^\nu)~
+\zeta \, \bar{\psi}_\nu \, \gamma^\nu \gamma^\mu \gamma_\rho \, \psi^\rho 
\ee
The same structure also appears in the cubic electromagnetic interaction vertex $ i\,q\,\Gamma$ of the theory:
\be
\label{eq:vertex}
\Gamma^\mu_{ \nu\,\rho} = \frac{\delta L}{\delta A_\mu ~\delta \bar{\psi}^\nu  ~ \delta \psi^\rho} = 
\gamma^\mu \, \delta^\nu_{~\rho}
+ \xi \, (\delta^\mu_\rho \, \gamma^\nu +  \delta^{\mu}_{\nu} \, \gamma_{\rho })
+ \zeta \gamma^\nu \, \gamma^\mu\, \gamma_{\rho }
\ee
Further, it is important to ensure that the vertex leads to an amplitude which vanishes for longitudinal photons. What we need is a more-or-less standard Ward identity for the inverse of the fermion propagator:
\be
\label{eq:ward}
i\, k_\mu \, \Gamma^{\mu\nu}_{~\rho}(p+k,p) = S^{\nu}_{~\rho}(p+k)^{ -1} - S^{\nu}_{~\rho}(p)^{ -1}
\ee
and is satisfied, at least at the tree level, due to conservation of current together with the minimal coupling principle. Here, the inverse of the propagator can be taken as the kinetic operator itself, but in Section \ref{sec:prop} we construct the complete propagator and discuss the requirements imposed on it by the Ward identity.

The spatial integral of the zero component of the current is electric charge:
\[
Q = \int j^0 = - \int \bar{\psi}_\nu \, \gamma^0 \, \psi^\nu + 
\xi \, (\bar{\psi}_\nu \, \gamma^\nu \psi^0 - \bar{\psi}^0 \gamma_\nu \, \psi^\nu ) +
\zeta \, \bar{\psi}_\nu \, \gamma^\nu \gamma^0 \gamma_\rho \, \psi^\rho~,
\]
The value of the charge is the same for all 8 solutions (\eqref{eq:e4}, \eqref{eq:p4}) of total spin 3/2, and is of the opposite sign for all 4 solutions ( \eqref{eq:e2},\eqref{eq:p2}) of total spin 1/2:
\beas
\bar{u}_4 \, \Gamma^0 \, u_4 &=& \bar{v}_4 \, \Gamma^0 \, v_4 = 1\nn
\bar{u}_2 \, \Gamma^0 \, u_2 &=& \bar{v}_2 \, \Gamma^0 \, v_2 = - 2 \, ( 3\,z - 1)^2
\eeas
One may wish to normalize the spin 1/2 solutions differently, such as to make the charge equal to minus one in the rest frame. 


Sudarshan and Johnson \cite{Johnson:1960vt} and several other authors require that the charge matrix, $\gamma_0 \Gamma_0$, be
positive definite,  whereas the expression above is manifestly not. Positivity of charge was first required by Pauli 
\cite{Pauli1941}, but not because of the idea that $j^0$ might be identified as the probability density - rather in connection with his new spin-statistics result. Pauli noted that, classically, the charge is of the same and energy of the opposite sign for the positive and negative frequency solutions, and this necessitates the quantization with fermi statistics and anti-commuting operators resulting in the quantum theory with positive energy of the physical particles and opposite charge for the particle and anti-particle.

 The electric charge in the Dirac theory is equal to the difference in the number of positrons and electrons $Q = \hat{N}_v -  \hat{N}_u$, while in the case at hand it is, after second quantization (Section \ref{sec:field})
$$
Q = (\hat{N}^{3/2}_u -  \hat{N}^{3/2}_v) - (\hat{N}^{1/2}_u -  \hat{N}^{1/2}_v)~.
$$
 Here, we mention that energy, classically, turns out to be negative for the positive frequency spin 1/2 solutions and necessitates the choice of the opposite vacuum to the usual.  The positive frequency solutions for the spin 3/2 and spin 1/2 particle have the same parity for $M>0$, such that the net result is that particle excitations of the true vacuum which correspond to the modes of the same sign of frequency have also the same charge and the same parity. 
 
Therefore, we must insist that the essence of Pauli's requirement for multi-component theories is that sign and value of the charge  for all helicities must be equal for each particle-antiparticle component \emph{separately}. This is because the choice of the physical vacuum can be done separately for the different particle components - the detailed considerations set out in the subsequent sections bear this out.

These considerations are enough to take care of only the energy positivity, but do not resolve by themselves the very serious problems with quantization that the Sudarshan and Johnson result implies, this was confirmed by Munczek \cite{Munczek} for the most general Rarita-Schwinger-type lagrangian.
If, instead, the spin 1/2 component is projected out by means of imposing $\gamma_\nu\, \psi^\nu=0$, then the charge can be made positive but this leads to the intractable Velo-Zwanziger problem. 

 We postpone until Section \ref{sec:field} a detailed discussion of the canonical formalism, and whether the non-positiveness of charge prevents consistent  quantization.
In the next section we look at the wavefront propagation in the presence of non-zero electromagnetic field which is coupled to the particle as described above. 

\section{Superluminal wavefronts at $z=1/3$}
\label{sec:super}

It has been long known that as soon as an electromagnetic field is turned on, there appear wavefronts propagating faster than the speed of light \cite{Velo:1969bt,Velo:1970ur}, also confirmed in \cite{Hortacsu:1974bm}. 
It was noticed by Ranada and Sierra \cite{Ranada:1980yx} that the problem can be cured if a spin 1/2 particle is allowed to propagate along with the spin 3/2 particle,
and we shall reconfirm their conclusion by investigating this phenomenon as a function of the parameter $z$.
Although it is possible to determine the dispersion relations, i.e. the dependence of $\omega$ on $\vec{p}^2$ by directly examining the kinetic operator $D$, we will follow tradition (for example \cite{Niederle:2004bw}) and examine the issue only in the relevant spin 1/2 sector by writing down the equations in the presence of the electromagnetic field, contracted with $\gamma^\mu$ and $\pi^\mu$ respectively:
\begin{eqnarray*}
\label{eq:vz1}
2\,\pi\cdot \psi +  2\, (3\,z-1) \,\gamma \cdot \pi \, \gamma \cdot \psi + m \, \gamma \cdot \psi &=&0\nn
( 2\,z\, \gamma \cdot \pi -m ) \, \pi\cdot\psi + [ 2\,z\,(3\,z-1) \,\pi^2 + z\,m\, \gamma \cdot \pi + (2\,z-1) \,\sigma \cdot F ] \,\gamma \cdot \psi+ i \, F_{\mu\nu} \, \gamma^\nu \,  \psi^\mu &=& 0~.
\end{eqnarray*}
These are special cases of the equations (10) and (11) of Ranada and Sierra \cite{Ranada:1980yx}.

The two equations can be combined 
into 
\[
\left[ 2\, (3\,z-1)\,  \gamma \cdot \pi + (4\,z-1)\, m +  \frac{(2 \,z - 1)\, (4\, z - 3)}{m\,(4\, z-1)}  \,\sigma \cdot F \right] \, \gamma \cdot \psi
- \frac{2\,z-1 }{4\, z-1}\, {2i \over m} \, \gamma \cdot {F} \cdot \psi = 0
\]

Only the first term here involves the time derivative. The analysis of Velo and Zwanziger applies to the original Rarita-Schwinger  case with $z=1/3$, whereby the coefficient of the first term vanishes and the equation should be regarded as a secondary constraint. This constraint is $\gamma\cdot\psi=0$ for vanishing external field, but for non-vanishing field one can look for superluminal wavefronts in the eikonal approximation $\psi^\mu = \epsilon^\mu \, \exp( i \, \tau \, n_\mu \, x^\mu)$ at $\tau \rightarrow \infty$. Indeed, at $z=1/3$
one can always find Lorentz frames with negative norm polarization
$\epsilon^\mu = (1,0,0,0)$ as a solution, meaning ultimately that there are superluminal solutions in every frame.
For $z\neq 1/3$ the first term dominates in the limit $\tau \rightarrow \infty$, forcing transversality $ n_\mu \epsilon^\mu =0$ thus there are no superluminal wavefronts in the general case. In fact, when $z\neq 1/3$ hyperbolicity no longer depends on the local value of the electromagnetic field, as it should not, according to \cite{Ranada:1980xp}.

The conclusion is that the appearance of superluminal waves is a direct and unavoidable consequence of the aesthetic preference to deal only with the spin 3/2 component, getting rid of both longitudinal and transverse spin 1/2 components by imposing both $ p_\mu \, \psi^\mu=0$ and $\gamma_\mu \, \psi^\mu=0$. What we find is that when the electromagnetic field is turned on, the tachyonic nature of the longitudinal spin 1/2 component reasserts itself in the prohibited regime $z\rightarrow 1/3$. Instead, when the latter constraint is relaxed away from the point $z=1/3$, then hyperbolicity is assured.

The author wishes to thank the anonymous referee for pointing out the paper by Ranada and Sierra \cite{Ranada:1980yx}, in which it was 
first proposed to allow one of the spin 1/2 components to propagate as a solution to the acausal propagation problem. Indeed, a survey of the early literature shows that the fact that the multi-particle equations do not generally suffer problems after coupling to electromagnetic fields was known to the experts in the field as early as in the beginning of the 1950's \cite{bhabha:1952,Johnson:1960vt}. More recent work also points in this direction \cite{Savvidy:2010aj, Rahman:2011ik}. The Ranada and Sierra proposal is valuable in the sense that we are now in the position to argue that the negative norm, longitudinal components can be projected out without destroying causality of the wave equations. The remaining question 
is whether the multi-particle equations in which only the transverse physical degrees of freedom propagate may also be consistently quantized. We will have the opportunity to reexamine the issue for the particular lagrangian \eqref{eq:lagr} in the next Sections \ref{sec:prop} and \ref{sec:field}.

\section{The Path Integral and the Feynman Propagator}
\label{sec:prop}

We quantize the theory using the fermionic path integral formalism, and later reconfirm and validate the results using the canonical formalism in the next section. The special, and unfamiliar feature of the theory is that it is necessary to integrate over a fermionic lagrange multiplier. There is no canonical conjugate to this field, yet every available authoritative source requires integrating over the Grassmann conjugates for consistency, therefore we do include the integration over the Grassmann conjugate $\psi^{\dagger}_0$ of the auxiliary field $\psi_0$. Fermionic theories do not, in any case, admit a true canonically conjugate field - if we introduce the ``canonical" momentum by
$\pi=\delta L / \delta \dot{\psi} = - i\,\psi^\dagger \gamma_0 \Gamma_0$ it does not contain a time derivative and therefore is not dynamically independent from the field itself; fermionic lagrangians are always singular from the canonical point of view. 
Usually one ignores this fact, and in the case of the Dirac field this happens to obtain the correct results, in our opinion because $\pi$ just happens to be $i\psi^\dagger$. More important is the fact that the definition of the conjugate field as $\pi=\delta L / \delta \dot{\psi}$ leads to intractable difficulties with quantization \cite{Johnson:1960vt, Munczek}.

We now demonstrate that the path integral over Grassmann conjugate fields $\psi$ and $\psi^\dagger$ results in an acceptable theory. The generating functional is defined as the following, 
\be
Z[\eta, \eta^\dagger] = \int  \mathcal{D}\psi_\mu ~ \mathcal{D} \psi_{\nu}^{\dagger} \,
 \exp \left[\, i \int d^4x \, [\,\mathcal{L}( \psi_\nu^{\dagger}, \psi_\mu) + \psi_\nu^{\dagger} \, \gamma_0 \, \eta^\nu + \eta_\mu^{\dagger} \, \gamma_0 \, \psi^\mu - i \, \epsilon \, \psi_\mu^{\dagger} \, \gamma_0 \, \psi^\mu] \right]
\ee
The fermionic path integral was defined rigorously by Berezin in 1971 \cite{Berezin:1971jf, Berezin:1966nc} as a generalization of ordinary integration over anti-commuting variables. When there is no Grassmann conjugate,  i.e. in the case of the Majorana field, the preferred method \cite{Bastianelli-Nieuwenhuizen} is to extend integration by pairing up Majorana components into complex spinors, such as to make both $\psi$ and $\psi^\dagger$ available also in that case. 

Above, as in the bosonic case, it is necessary to add a small imaginary part to the lagrangian. In the bosonic case, it can be justified in order to make the gaussian integral convergent. In the fermionic case the path integral is defined formally, but we need to add the term in any case,  as it is necessary to move the poles of the Feynman propagator off the real axis. The correct sign of the infinitesimal $\epsilon$ cannot be postulated \emph{a priori} in the fermionic case, and is justified by the correct locations and residues at the poles.

As usual, it is possible to integrate out the fields resulting in the master formula, 
\be
Z[\eta, \eta^{\dagger}] = Z_0 \, \exp \left[ - \frac 1 2 \int d^4 x \, d^4 y~  \eta_\mu^{\dagger}(x) \, \gamma_0 \, S^\mu_{\,\nu}(x-y) \,  \eta^\nu(y) \right]~.
\ee
Here, the Feynman propagator should be obtained by
\be
S^\mu_{\,\nu}(x) = i \, \int d^4 x ~ \e{ipx} ~ \left[   \frac{\delta L(p)} {\delta \psi_\mu \, \delta \bar{\psi}_\nu } \right]^{-1}
\ee
The operator $D^\mu_{~\nu} - m \, \Theta^\mu_{~\nu} + i \, \epsilon$ is non-degenerate except for the points in momentum space where the two physical particles go on-shell, thus there should be no difficulty in constructing the propagator.
In order to guess the inverse matrix $S^\mu_{~\nu}(p) = - i \,(D^\mu_{~\nu} - m \, \Theta^\mu_{~\nu}+  i \, \epsilon)^{-1}$ it is then necessary to first obtain the correct form of the spin-sum expressions that go into the numerator. To do that, we need to express the projection operators (spin sums) as Lorentz covariant tensor expressions. The ingredients are the following set of spinor-vector projection operators:
\begin{eqnarray}
\label{eq:proj}
\Pi_3 &=& \delta^\mu_{~\nu} - \tfrac 1 3 \, \gamma^\mu \, \gamma_\nu - 
                     \tfrac 1 {3p^2} \, (\slashed{p} \, \gamma^\mu \, p_\nu + p^\mu \, \gamma_\nu \, \slashed{p}) \nn
\Pi_{11} &=&  \tfrac 1 3 \, \gamma^\mu \, \gamma_\nu - \tfrac 1 {p^2} \,  p^\mu \, p_\nu +
                     \tfrac 1 {3p^2} \, ( \slashed{p} \, \gamma^\mu \, p_\nu + p^\mu \, \gamma_\nu \, \slashed{p}) \nn
 \Pi_{22} &=& \tfrac 1 {p^2} \, p^\mu \, p_\nu \nn
 \Pi_{21} &=&  \tfrac 1 { p^2} \,  ( p^\mu \, p_\nu  - \slashed{p} \, \gamma^\mu \, p_\nu) \nn
 \Pi_{12} &=&  \tfrac 1 { p^2} \,  ( \slashed{p} \, p^\mu \, \gamma_\nu  - p^\mu \, p_\nu )~.
\end{eqnarray}
We suppress the indexes, which are always a pair of vector-spinor indices, such as $\Pi^{\mu\alpha}_{\nu\beta}$, since in practice it is more convenient to think of them as 16x16 matrices acting on 16 component vector-spinors.

This set of projection operators has the following meaning \cite{neuw, Benmerrouche:1989uc, Pascalutsa:1995vx}. $\Pi_3$ projects to the pure transverse spin 3/2 subspace, $\Pi_{22}$ to the longitudinal spin 1/2 subspace, and finally $\Pi_{11}$ to the transverse  spin 1/2 subspace. Lorentz invariance alone does not dictate that the two spin 1/2 subspaces will not mix in the lagrangian or the equations of motion. This necessitates the use of the remaining two operators, $\Pi_{12}$ and $\Pi_{21}$ which respectively take vectors from longitudinal to the transverse space and the reverse.


Further, in order to restrict to the positive and negative energy subspaces separately, 
we define two additional projection operators for projecting to the parity-odd and -even eigenspaces:
\begin{eqnarray}
\label{eq:parity}
 \Pi^\pm(\Omega) &=& \frac 1 2 ~ \left({\openone} \mp \mathbb{P}  \right)\nn
\mathbb{P} &=& \left(-\delta^\mu_{~\nu} + 2 \, \frac { p^\mu \, p_\nu} {p^2}\right)  \otimes  \frac {\slashed{p}} {\Omega}\nn
\mathrm{where}~ \Omega &=& \sqrt{ p^2},\,m, \, M.
\end{eqnarray}
It will be necessary to specify the value of $\Omega$ depending on the intended use, i.e. in the case of the propagators and spin sums it is more convenient to work with the on-shell value of $\Omega$, equal to mass. The parity operator $\mathbb{P}$ was constructed as the covariant version of the fixed lab-frame parity operator $\eta \otimes \gamma_0$, which is incidentally the one we normally use to apply the Dirac bar operation (see below).

We have checked explicitly that the space spanned by positive and negative energy solutions coincides
with the subspaces projected to by our projectors:
\begin{eqnarray}
\label{eq:spinsum}
 \sum_{s=-3/2}^{3/2} u_4(\mathbf{p}, \,s) ~\, \bar{u}_4(\mathbf{p}, \,s) &=&   {-} \Pi^+\, \Pi_3  \nn
 \sum_{s=-3/2}^{3/2} v_4(\mathbf{p}, \,s) ~\, \bar{v}_4(\mathbf{p}, \,s) &=&   \Pi^- \, \Pi_3  \nn
 \sum_{s=-1/2}^{1/2} u_2(\mathbf{p}, \,s) ~\, \bar{u}_2(\mathbf{p}, \,s)   &=& {-} \Pi^+\,\Pi_{11}   \nn
 \sum_{s=-1/2}^{1/2} v_2(\mathbf{p}, \,s) ~\, \bar{v}_2(\mathbf{p}, \,s)   &=&  \Pi^- \, \Pi_{11} ~.
\end{eqnarray}
On the contrary, the subspace spanned by $\Pi_{22}$, is not collinear with the \emph{nullspace} of the kinetic operator  $D^\mu_{~\nu} - m \, \Theta^\mu_{~\nu}$ for any value of momenta. Thus, longitudinal negative norm field modes
do not propagate. We shall verify this by direct inspection of the propagator, below.

\begin{figure}[ltbp]
\begin{center}
\includegraphics[scale=0.8]{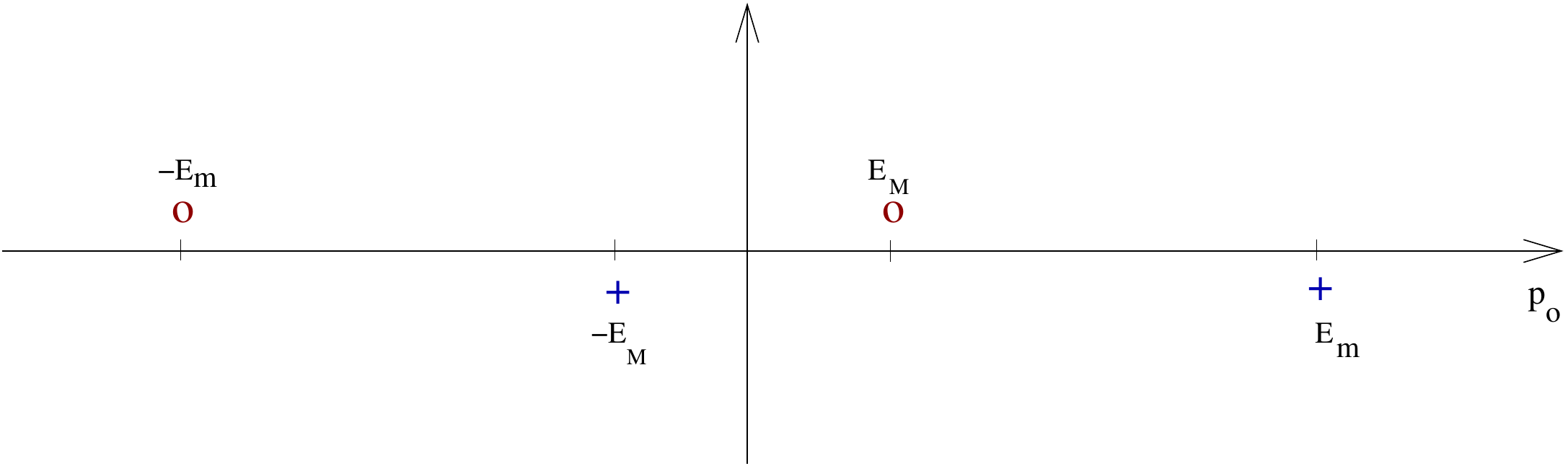}
\caption{{\small The residue at the poles lying above the real axis is negative (red o's) and those lying below the real axis is positive (blue +'s). Extraction of the poles is best understood as an integral over $p_0$ on real axis from $-\infty$ to $+\infty$, with the contour closed at infinity either above or below, depending on whether 
a particle incoming from $t=-\infty$ or outgoing to $t=+\infty$ is desired.
The subsequent multiplication by the wavefunction of the external leg, as per LSZ, results in a positive forward amplitude $\langle p | p \rangle $ in all four cases.
}\label{fig:poles}}
\end{center}
\end{figure}

 
With these ingredients, our result for the propagator $S^\mu_{~\nu}(p) = - i\,(D^\mu_{~\nu} - m \, \Theta^\mu_{~\nu} +i\, \epsilon)^{-1}$ is:
\begin{eqnarray}
\label{eq:prop}
-i\,S^\mu_{~\nu}(p) &=& \frac{- 2 \, m \, \Pi^+{(m)\, \Pi_3}  }{p^2-m^2+i\,\epsilon}  
- \frac{-2\,M \, \Pi^+{(M)} \, \Pi_{11} }{p^2-M^2 - i\,\epsilon} \, \frac {1} {2\,(3\,z-1)^2} \nonumber\\
&+& \, \tfrac{3}{2 \,(M +2 \, m)} 
\left[ \vspace{30pt} \Pi_{22} - (\Pi_{21}+\Pi_{12})\,/B +  \Pi_{11}\,3/B^2 \right] ~~,\nonumber\\
B &=& \frac{3\,m}{2 \, M +m} ~~.
\end{eqnarray}
The locations of the poles are of paramount importance and were obtained by inspecting the solutions of the characteristic polynomial in the presence of the infinitesimal regulator:
\[
\det (D^\mu_{~\nu} - m \, \Theta^\mu_{~\nu} +i\, \epsilon) = 0~.
\]
The solutions of this equation, with correct multiplicities (4,4,2,2), in the rest frame lie at 
\[
m - i\,\epsilon, ~~ 
- m +i\,\epsilon, ~~
M + i\,\epsilon \, \kappa + O(\epsilon^2), ~~
-M - i\,\epsilon \, \kappa + O(\epsilon^2)~~.
\]
where $M = { 1 \over 2 (3z-1) }$ and $\kappa = { 1 \over 2 (3z-1)^2 } > 0 $. Thus the poles are located in the usual way for the spin 3/2 particle and anti-particle but in the exotic position for the spin 1/2 particle and anti-particle, so long as $M>0$. Specifically, the positive frequency pole lies above the real axis and the negative  frequency pole lies below the real axis (see Fig. \ref{fig:poles}).  Inspection also shows, that the residue is negative at the positive frequency spin 1/2  pole and positive at the negative frequency pole (we use the physicist's definition of residue as $\Res f(z) = \ointclockwise f(z) dz$, so that $\Res i/z = 2\pi $). 
One could summarize this by saying that the spin 1/2 propagation amplitude is such that the positive frequency solutions contribute to the advanced potential, while the negative frequency solutions contribute to the retarded potential; the retarded potential nevertheless comes with positive sign as it must. Indeed, classically the spin 1/2 \emph{negative} frequency modes have positive energy - therefore the situation is appropriate in that only these contribute to the retarded potential and with positive sign. In the path-intergal formulation this is precisely the physical postulate to deal with negative energy modes, i.e. the negative energy modes should contribute to the advanced potential (propagating back in time) while in the canonical quantization energy is made positive by choosing the appropriate vacuum. As we shall see in the next Section \ref{sec:field} and Appendix \ref{sec:app2}, the two approaches are completely consistent with each other; the given pattern of pole positions and residues readily follows in the canonical formalism once the correct choice of the vacuum is made.

\begin{proposition}
The propagation function must be of half-retarded minus half-advanced form:
poles lying below the real axis propagate forward in time and must have positive residue, poles lying above the real axis propagate backward in time and must have negative residue. 
\end{proposition}

The residues as they appear here have been anticipated in the extant literature, see \cite{bhabha:1952, Johnson:1960vt, Munczek, Fukuyama}, but 
the location of the poles has not been examined to the author's knowledge. Together, we argue, the unusual pole locations, together with the ``wrong" sign of the residue results in  a unitary theory with positive forward amplitude $ \langle p | p \rangle$. In order to calculate the forward amplitude it is best to make use of the  Lehmann-Symanzik-Zimmerman (LSZ) theorem. 

Generally speaking, the LSZ theorem establishes a relation between n-point functions and n-particle scattering S-matrix elements. Specifically, for the two-point function it generally follows that the positivity of the probability $\bra{p} {p} \rangle$ requires the residue to be positive at the pole corresponding to the time-forward propagating particle (and negative at the time-backward propagating anti-particle pole). The multi-component fermionic propagator is unusual in that in the spin 1/2 sector its positive frequency pole lies above the real axis instead of below, as is usual. The position of the pole above the axis implies that the corresponding modes propagate backward in time. It is then natural that the residue at such pole is found to be negative. 

We now go back to our propagator \eqref{eq:prop}, and
 further elucidate the meaning of this 
expression and compare to those appearing previously in the literature \cite{neuw,Benmerrouche:1989uc,Pascalutsa:1995vx}.

First, in order to get the correct decomposition between positive and negative energy propagating modes we must use parity projection operators which correctly take into account the vector-spinor nature of the Rarita-Schwinger wavefunction. The operators as appearing above in \eqref{eq:prop} have mass substituted in the denominator, rather than $\sqrt{p^2}$, just like in Dirac theory. Existing literature \cite{neuw,Benmerrouche:1989uc,Pascalutsa:1995vx} did not make use of the vector-spinor parity operator in the propagator, simply because it is unnecessary in the spin 3/2 term and can be safely replaced by its Dirac spinor equivalent $\Pi_+ = (1+\slashed{p}/m)/2$; in the given version of the theory, since all solutions are transverse $p_\mu \psi^\mu = 0$, one can make the same simplification also for the spin 1/2 term.

Second, although the separate terms appear to be singular on the lightcone $p^2=0$, the total is non-singular.

Third, the purely longitudinal mode corresponding to $\Pi_{22}$ contributes without ever going on-shell, with an amplitude independent of momentum in \eqref{eq:prop}. This does not lead to disastrous consequences, even though the interaction vertex  allows the coupling to such purely longitudinal states. 
The S-matrix must vanish with longitudinal polarization fermion states; here they never propagate on-shell and at most contribute to the off-shell Green's functions but not the S-matrix. This is seen most easily in the LSZ formalism: the S-matrix elements are read off from the poles in the unamputated Greens function; appropriately, propagator above does not contain poles corresponding to the longitudinal fermion states. 

On the other hand, amplitudes with photon longitudinal states will vanish if the Ward identity is satisfied in the form
\[
S^{\kappa}_{~\nu}(p+k) \, i\, k_\mu\,\Gamma^{\mu\nu}_{~\rho}(p+k,p) \, S^\rho_{~\sigma}(p) =   S^{\kappa}_{~\sigma}(p) - S^{\kappa}_{~\sigma}(p+k)~~,
\]
this can be now checked directly, and moreover demonstrates that the non-pole terms, $\Pi_{22}+...$, cannot be dropped and contribute essentially to the fulfillment of the Ward identity.



Finally, the limit of the second part as $M \rightarrow \infty$ is nonzero. Thus, even as the spin 1/2 component becomes infinitely massive, its influence does not go away even in the limit. The presence of those terms led Pascalutsa and Scholten \cite{Pascalutsa:1995vx} to comment that the extra pieces must be due to some high-lying resonances, with the momentum dependence of the propagator being washed out due to the high mass of the physical particle. 
In our theory, the spin 1/2 component is interpreted as the progenitor, and most of the allowed region for its mass lies below that of the spin 3/2 component; certainly the limit $M \rightarrow \infty$ is illegitimate. 
On the contrary, the case $M << m$ is the important physical limit to consider in extensions of the  Standard Model, if one wishes to identify some known particle as the spin 1/2 component of the multiplet and argue that the spin 3/2 component is much heavier.

This completes the construction of a propagator with poles corresponding to physically propagating particles and only them, and usable for calculations of scattering amplitudes. 

\section{Canonical Quantization}
\label{sec:field}

The quantum field can be built up as the object formally satisfying the relativistic wave equation $ [ D(p) - m\, \Theta ] \, \psi(x,t) =0$, with mode coefficients treated as operators, as follows:
\begin{eqnarray}
\label{eq:qft}
\psi^\mu (\mathbf{x}, t) &=&
 \int \frac {d^3 \mathbf{p} } {8\pi^3} 
\left(
\sum_{s=-3/2}^{3/2} \sqrt{ \frac{m}{ { E_{\mathbf{p}}(m)}  } } 
\left[ a( \mathbf{p} ,  s) \, u^\mu_4(\mathbf{p}, \,s) \, \phi^+_m(\mathbf{x},t)  + 
                          {b}^{\dagger}( \mathbf{p},  s) \, v^\mu_4 ( -\mathbf{p} ,s) \, \phi^-_m(\mathbf{x}, t)\right] \right. \nonumber\\
&+& \left. \sum_{s=-1/2}^{1/2}\sqrt{ \frac{Z\,M}{ { E_{\mathbf{p}}(M)}  } } \,  
\left[ c( \mathbf{p},  s) \, u^\mu_2 ( \mathbf{p}, s)\, \phi^+_M(\mathbf{x},t) +
                          {d}^{\dagger}( \mathbf{p},  s) \, v^\mu_2 ( -\mathbf{p} , s) \, \phi^-_M(\mathbf{x},t) \right]
 \right) ~,\nonumber\\
 E_{\mathbf{p}}(\Omega) &=& \sqrt{ \mathbf{p}^2 + \Omega^2} ~~,~~~
 \phi^\pm_\Omega(\mathbf{x}, t) = \exp(  \mp i\, E_\Omega \, t \pm  i \, \mathbf{p}\, \mathbf{x})~, ~~Z = \frac{1}{2\,(3\,z-1)^2}~,
\end{eqnarray}
with the standard anti-commutation relations for all operators:
\beas
\{ \,a(\mathbf{p},  r) \,, \,a^{\dagger}(\mathbf{q} , s) \,\} = \{ \,b(\mathbf{p}, r) \,, \, b^{\dagger}(\mathbf{q}, s)  \,\} &=& 
(2\pi)^3 \;\delta^{(3)} (\mathbf{p} - \mathbf{q}) \, \delta_{rs} \nn
\{ \,a(\mathbf{p},  r) \,, \,a^{}(\mathbf{q} , s) \,\} = \{ \,b(\mathbf{p}, r) \,, \, b^{}(\mathbf{q}, s)  \,\} &=& 0\nn
\{ \,a^{\dagger}(\mathbf{p},  r) \,, \,a^{\dagger}(\mathbf{q} , s) \,\} = \{ \,b^{\dagger}(\mathbf{p}, r) \,, \, b^{\dagger}(\mathbf{q}, s)  \,\} &=& 0\nn
\{ \,c(\mathbf{p},  r) \,, \,c^{\dagger}(\mathbf{q} , s) \,\} = \{ \,d(\mathbf{p}, r) \,, \, d^{\dagger}(\mathbf{q}, s)  \,\} &=& 
(2\pi)^3 \,  \delta^{(3)}\,(\mathbf{p} - \mathbf{q}) \, \delta_{rs} \nn
\{ \,c(\mathbf{p},  r) \,, \,c^{}(\mathbf{q} , s) \,\} = \{ \,d(\mathbf{p}, r) \,, \, d^{}(\mathbf{q}, s)  \,\} &=& 0\nn
\{ \,c^{\dagger}(\mathbf{p},  r) \,, \,c^{\dagger}(\mathbf{q} , s) \,\} = \{ \,d^{\dagger}(\mathbf{p}, r) \,, \, d^{\dagger}(\mathbf{q}, s)  \,\} &=& 0~.
\eeas

At this point, choice of which are labeled as the annihilation and which as the creation operators is arbitrary. The choice is fixed once it is decided
to keep the standard convention that the Hamiltonian must be equal to 
\[
H= \frac 1 {8\pi^3}\, \int d^3\mathbf{p} \sum_{s}  E_{\mathbf{p}}(m)  ~(a^\dagger \, a + b^\dagger \, b) +  E_{\mathbf{p}}(M) ~(c^\dagger \, c + d^\dagger \, d)~,
\]
such that the  ``creation" (daggered) operators always have negative frequency and ``annihilation"  operators always have positive frequency and  as in \eqref{eq:qft}; this is also true of  $\psi^\dagger$.  

By construction, with this hamiltonian the field satisfies the Heisenberg equation,  regardless of the choice of vacuum:
\be
i\,\tfrac{\partial}{\partial_t}\psi(x,t) = [H, \psi(x,t)]~~.
\ee

The choice of the vacuum is not determined yet. This is because in quantum field theory, unlike quantum mechanics, the frequency is not the same as energy. Instead,
for the theory to be physical it is necessary that the energy  be positive for all physical excitations. Energy should be defined as the expectation value of the energy-momentum tensor 
\[
T^{\mu}_{\nu} =  \frac{ \delta\mathcal{L} } { \delta \partial_\mu \psi }\, \partial_\nu \psi - \delta^{\mu}_{\nu} \mathcal{L} ~.
\]
The zero component happens to be the Legendre transform of the lagrangian:
\be
T_{00} = \int \left[ \frac{ \delta\mathcal{L} } { \delta \dot \psi }\, \dot \psi - \mathcal{L} \right]~d^3 \mathbf{x} 
=  - \int  \bar{\psi}_\nu \, \left[  \vec{\Gamma}^\nu_\rho \, \vec{p} + m \, \Theta^\nu_\rho \right] \, \psi^\rho ~d^3 \mathbf{x} ~.
\ee
Note the formal analogy to the energy in the Dirac case. By  equations of motion this can be shown to coincide with 
$- \bar{\psi}_\nu \, \Gamma^{0\nu}_{~\rho} \, p_0 \, \psi^\rho$, which allows to claim the equivalence of the Noether and GR ($T_{00} = \delta L / \delta g_{00}$ ) definitions of the energy. 

The field can now be substituted to get:
\[
T_{00} =  \int \frac {d^3 \mathbf{p} } {8\pi^3} ~  \left[E_{\mathbf{p}}(m)\, (\hat{N}^{3/2}_u + \hat{N}^{3/2}_v) - E_{\mathbf{p}}(M) \,  (\hat{N}^{1/2}_u + \hat{N}^{1/2}_v)\, \right]~,
\]
where 
\begin{eqnarray*}
\hat{N}^{3/2}_u = \sum_{s=-3/2}^{3/2} a^{\dagger}(\mathbf{p},  s) \, a(\mathbf{p},  s) & ~,~~~~~
\hat{N}^{3/2}_v = \displaystyle \sum_{s=-3/2}^{3/2} b^{\dagger}(\mathbf{p},  s) \, b(\mathbf{p},  s)  ~\nn
\hat{N}^{1/2}_u = \sum_{s=-1/2}^{1/2} c^{\dagger}(\mathbf{p},  s) \, c(\mathbf{p},  s) & ~,~~~~~ 
\hat{N}^{1/2}_v = \displaystyle \sum_{s=-1/2}^{1/2} d^{\dagger}(\mathbf{p},  s) \, d(\mathbf{p},  s)  ~.\\
\end{eqnarray*}
The charge $Q =  -\bar{\psi}_\nu \, \Gamma^{0\nu}_{\, \rho} \, \psi^\rho $ is similarly found to be
\[
Q = \int  \frac {d^3 \mathbf{p} } {8\pi^3}  ~ \left[(\hat{N}^{3/2}_u -  \hat{N}^{3/2}_v) - (\hat{N}^{1/2}_u -  \hat{N}^{1/2}_v) \right]~~.
\]
Both charge and energy contain the relative minus sign between the spin 3/2 and spin 1/2 parts, but there is nothing particularly wrong here. Pauli famously demanded that in the c-number theory the charge must be positive definite in order for the physical energy to be positive after choosing the Dirac vacuum. In the resulting q-number theory, charge will turn out to be opposite for the particle and anti-particle. We are already working with q-number mode operators with appropriate anti-commutation relations, so it only remains to ensure the positivity of the energy for excited states. As we see, it is possible to satisfy Pauli's requirement because we can choose the vacuum separately for the different spin sectors.
Thus, for the vacuum $\ket{\Omega}$ to be the state of lowest energy $E = \bra{\Omega} T_{00} \ket{\Omega} $, we must choose the highest-weight vacuum for the spin 1/2 sector:
\bea
a\,(\mathbf{p}, s) \,\ket{\Omega} &=& b\,(\mathbf{p},  s)  \,\ket{\Omega} = 0~~, ~~~s=-3/2,-1/2,+1/2,+3/2~~,\nonumber\\
c^\dagger(\mathbf{p},  s) \,\ket{\Omega} &=& d^\dagger(\mathbf{p}, s)  \,\ket{\Omega} = 0~~, ~~~s=~-1/2,+1/2~.
\eea
or in other words, the physical vacuum is $ \ket{\Omega} = \ket{0}_{3/2} \otimes \ket{1}_{1/2}$. Pauli's requirement is satisfied in the sense that energy is positive and charge has the opposite sign for the physical particle and anti-particle. 



The full, relativistic anti-commutator of the field with its conjugate is:
 \bea
\{ \, \psi^\mu_a ( \mathbf{x}, t) \, , \, \psi^\dagger_{\nu b} ( \mathbf{y}, t^\prime) \, \}  = 
 {} 2\,m\,\Pi_3 \, \left(-\Pi^+\,    \Delta_+^m \,( \mathbf{x} - \mathbf{y}, t - t^\prime)
+ \Pi^- \,   \Delta_+^m \,(  \mathbf{y} - \mathbf{x}, t^\prime - t )\right) \nonumber\\
 {} +  2\,M\, Z \, \Pi_{11} \, \left( - \Pi^+\,  \Delta_+^M ( \mathbf{x} - \mathbf{y}, t -t ^\prime)
 + \Pi^- \,    \Delta_+^M (  \mathbf{y} - \mathbf{x}, t^\prime - t ) \right) \, \gamma_0
\eea
or
\bea
\{ \, \psi^\mu_a ( x) \, , \, \psi^\dagger_{\nu b} (y ) \, \}  = 
 {-} 2\,m\,\Pi_3 \, \Pi^+\,  (i\partial_x)\left[  \Delta_+^m \,( x -y ) - \Delta_+^m \,( y - x  ) \right] \nonumber\\
 {-}  2\,M\, Z \, \Pi_{11} \, \Pi^+\, (i\partial_x)\left[  \Delta_+^M \,( x -y ) - \Delta_+^M \,( y - x  ) \right] \, \gamma_0
\eea
where we introduced the usual massive scalar retarded propagation function as
\[
\Delta^\Omega_+(\mathbf{x},t) = \frac 1 {8\pi^3} ~ \int \frac {d^3 \mathbf{p} } { {2\, E_{\mathbf{p}}(\Omega)}}
~\e{ i \, \mathbf{p}\, \mathbf{x} - i\,E_{\mathbf{p}}(\Omega)\,t}~.
\]
This function is even in coordinates at space-like separation, while the parity projectors $\Pi^\pm$ are exchanged under reflection.
The net result is that the field anti-commutator vanishes outside the lightcone - this is enough to assure causality, namely that the commutator of any observables (usually quadratic in fields) is zero outside the lightcone.
 
Finally, we are now in the position to reduce this result to the equal-time field anti-commutators. 
Making use of the identity $ 2m [ - \tfrac{m + \slashed{p} }{2m} +  \tfrac{ m - \slashed{p}^\dagger }{2m}] \gamma_0= -2E_p(m)$, the right-hand side of the equal-time field anti-commutator is positive definite 
\begin{eqnarray}
\label{eq:comm}
\{ \, \psi^\mu_a ( \mathbf{x}) \, , \, \psi^{\,\dagger}_{\nu b} ( \mathbf{y}) \, \}  &=& 
%
\Pi_3 (i\partial_{x_\mu}) \, \delta(\mathbf{x} - \mathbf{y} )  +
 Z \, \Pi_{11} (i\partial_{x_\mu}) \, \delta(\mathbf{x} - \mathbf{y} )\nn
\{ \, \psi^\mu_a ( \mathbf{x}) \, , \, \psi_{\nu b} ( \mathbf{y}) \, \} &=& 0\nn
\{ \, \psi^{\mu\,\dagger}_a ( \mathbf{x}) \, , \, \psi^{\,\dagger}_{\nu b} ( \mathbf{y}) \, \} &=& 0 ~.
\end{eqnarray}
The right-hand-side is explicitly positive definite because $\Pi_3$ and $\Pi_{11} $ are positive definite matrices with unit eigenvalues. 
Note that $i\partial_0$ is understood to act on-shell, $i\partial_0 = \sqrt{m^2 - \partial^2_\mathbf{x}}$. 
The result is further simplified if the two contributing fields have equal mass, $M=m$, and the field is normalized to $Z=1$. Then indeed 
\be
\{ \, \psi^\mu_a ( \mathbf{x}) \, , \, \psi^{\,\dagger}_{\nu b} ( \mathbf{y}) \, \}  = 
 \delta_{ab} \, \left(\delta^\mu_{\,\nu} + \frac {\partial^\mu \partial_\nu} {m^2} \right) \, \delta(\mathbf{x} - \mathbf{y} ) ~~,
\ee
where it is made apparent that there is no contribution from any longitudinal parts. 

We already argued in the section on path-integral quantization that the conjugate field should be taken as $\psi^\dagger$, and not $\pi =  \tfrac{\delta L}{\delta \dot{\psi} } = \psi^\dagger \, \gamma_0\,\Gamma^0$. When the latter choice is made, the result is an inconsistency in the quantization, first discussed by Sudarshan and Johnson  \cite{Johnson:1960vt} and later confirmed by Munczek \cite{Munczek} for the most general RS-type lagrangian. Sudarshan and Johnson in fact proved that all higher-spin multi-particle equations have this property, in the sense that no unitary transformation of the wavefunctions can result in a positive definite charge matrix $\gamma_0\,\Gamma^0$.

Further, we need to evaluate the practically very important vacuum expectation value of the time-ordered product, which is the two-point Greens function:
\be
D_F (x,y) = \langle \Omega | \, T \, \psi(x) \, \bar{\psi}(y) \, | \Omega \rangle
\ee
For definiteness, let us assume that $ x_0 > y_0$ so that we may proceed with the evaluation.
\bea
\label{eq:ret1}
D_F (x,y) &=& \frac{ m}{E_{\mathbf{p}} (m)  } ~ \int  \frac {d^3 \mathbf{p} } {(2\pi)^6} \,   \e{-ipx }\, 
  \, \e{ipy }~~ \nonumber\\
&&~~~~~\sum_s \langle \Omega \, | \, a( \mathbf{p},  s) \, a^\dagger(\mathbf{p},  s) \,  | \Omega \rangle ~
u_4( \mathbf{p},  s) \, \bar{u}_4( \mathbf{p},  s) \nonumber\\
&+& \frac{ Z\, M }{E_{\mathbf{p}} (M)  }~ \int  \frac {d^3 \mathbf{p} } {(2\pi)^6} \, \e{ipx } \, \e{-ipy }~~\nonumber\\
&&~~~~~ \sum_s \langle \Omega | \,  d^\dagger(\mathbf{p},  s) \, d(\mathbf{p},  s ) \,  | \Omega \rangle~
v_2( \mathbf{p}, s) \, \bar{v}_2( \mathbf{p},  s)
\eea
The crucial point here is that the retarded part gets contributions from the positive frequency spin 3/2 solutions, while spin 1/2 contributes its negative frequency modes due to the fact that the vacuum $\ket{\Omega}$ is the highest-weight state with respect to the spin 1/2 operators. Recall that  the energy is classically positive precisely for these, the spin 3/2 positive frequency and the spin 1/2 negative frequency modes. Appropriately, the retarded part of the Green's function propagates these forward in time, with positive coefficient.

This is then evaluated to
\bea
\label{eq:ret2}
D_F (x,y) = \int  &\frac{ d^3\mathbf{p}}{(2\pi)^3}~\frac 1 {{2\,E_{\mathbf{p}}} (m)  } \, \e{-ip(x-y) } \,  2\,m\,(-1) \Pi_{+}\Pi_3(\mathbf{p}) \nonumber\\
+
 \int & \frac{ d^3\mathbf{p}}{(2\pi)^3}~\frac 1 {{2\,E_{\mathbf{p}}} (M)  } \, \e{+ip(x-y) } \, 2 \,M \,Z\,\Pi_{-}\Pi_{11}(\mathbf{p}) 
\eea

Until now all momenta had been on-shell, $p_0 = E_{\mathbf{p}} > 0 $ everywhere. Next, we wish to obtain the two-point function as a four-dimensional Fourier transform of the momentum-space Feynman propagator. 
We insert the step function $\theta(x_0-y_0)$ in  the integral
representation:
\be
\theta(x) = \frac{1}{2\,\pi\,i\,}\, \int_{-\infty}^{+\infty} \frac{ds}{s-i\, \epsilon} ~\e{+isx}
= \frac{i}{2\,\pi} \, \int_{-\infty}^{+\infty} \frac{ds}{s + i\, \epsilon} ~ \e{-isx}
\ee
Using these integral representations, shifting $ p_0 = s + E_{\mathbf{p}}$ in the first term and $ p_0 = s - E_{\mathbf{p}}$ in the second, sending $p \to -p$ in the second term, then adding the advanced part which supplies the poles above the real axis, and combining the poles, we obtain
\bea
\label{eq:propC}
D_F (x,y) &=&  i ~ \int \frac {d^4p}{(2\pi)^4} ~ \e{-ip(x-y) } ~ \frac {2m (-1) \Pi_{+}\Pi_3(p)}{p^2-m^2+i\,\epsilon} 
+ \frac{2\,M \, \Pi^+  \Pi_{11}(p) }{p^2-M^2 - i\,\epsilon} \, \frac {1} {2\,(3\,z-1)^2}
\eea
where now $p_0$ is freely integrated over the real axis with the poles passed as indicated. The advanced poles are those coming from the spin 3/2 negative frequency modes and the spin 1/2 positive frequency modes. There is a relative minus sign between the two terms, as well as between their pole locations; the spin 3/2 poles have ordinary poles at $ \pm (E_p - i \, \epsilon)$ with ordinary residues, while the spin 1/2 poles are at $ \pm (E_p + i \, \epsilon)$ and their residues are of the opposite signs to the usual.
This is made possible solely due to the choice of the highest-weight vacuum (see  Appendix \ref{sec:app2}), which is itself necessitated by making sure of positivity of energy. We stress once again that the statement about residues which remains true in the present theory is that  residues must be positive at the poles lying below the real axis, and negative for poles lying above the real axis.
One may now check that the above expression coincides with that obtained by path integral \eqref{eq:prop} so far as the pole terms are concerned, including the pattern of residues and pole locations.


The entire calculation does not leave room for any ambiguity with the exception of the last step, and only in so far as how the replacement 
of the on-shell spin sums with off-shell expressions is made. Different prescriptions differ by non-pole terms, which may even be non-covariant. According to Weinberg, 
one may be able to ensure that the spin sum expressions are linear in $p_0$, otherwise the ambiguity may give rise to additional non-pole terms in the two-point function. This situation arises also in the theory of the massive spin 1 field, and is endemic to massive higher-spin fields. It is hard to see what principle may guide the determination of the non-pole terms in the canonical formalism, whereas the path-integral supplies the required terms with ease. One possibility is to demand that the propagator have asymptotic behavior as $1/p$ as $p \to \infty$. Another, even stronger requirement 
is the fulfillment of the Ward identity  \eqref{eq:ward}; the use of these is hard to justify in the canonical formalism, since Ward identities are usually obtained by inspecting Feynman diagrams in the interacting theory.

In the bosonic case the Kallen-Lehmann spectral representation demands that the scalar propagator have the behavior $1/p^2$ as $p \to \infty$. We present a derivation of the spectral representation for the fermionic case in Appendix \ref{sec:app2}. The generic arguments are not sufficient to make such strong claims about the asymptotic behavior - it is only guaranteed that the propagator will not \emph{grow} faster than $p^{2s-2}$. Indeed the individual terms in our propagator do grow linearly with momentum (s=3/2), but  \eqref{eq:prop} (like \eqref{eq:propC}) contains cancellations  between the linearly growing terms. The cancellations are indeed made possible by the relative minus sign between the spin 3/2 and spin 1/2 terms, and this minus sign can indeed be justified in the fermionic case generically, see the Appendix \ref{sec:app2}. The non-pole term in the total expression \eqref{eq:prop} (unlike \eqref{eq:propC}) cancels the remaining $O(1)$ behavior of the sum of the first two terms, 
which allow the total propagator to have the mild behavior,  $1/p$ as $p \to \infty$.

The considerations in this work are sufficient to claim that the free theory is unitary. The mild behavior of the propagator, together with the cancellations between the spin 3/2 and spin 1/2 sectors raise the hope that the interacting theory may evade the unitarity bounds for massive higher-spin particles and turn out to be not only unitary but also renormalizable. 

\section{Conclusion} 

The first part of the paper, Sections \ref{sec:rs}-\ref{sec:super}, deal the relativistic wave equation for the Rarita-Schwinger field which avoids the superluminality problem of Velo and Zwanziger. It is asserted, after Ranada and Sierra \cite{Ranada:1980yx} that the equations constructed such that only the physical spin 3/2 and spin 1/2 particles propagate, while the unphysical longitudinal spin 1/2 mode does not propagate, avoids the superluminality problem completely. In the second part, the quantum field theory of this field is investigated, with a view to resolve the apparent paradoxes \cite{Johnson:1960vt,Munczek} which this involves.

We have found that the four unusual features of the theory, namely
\begin{itemize}
\item ``wrong" sign in the lagrangian
\item difficulties of canonical quantization
\item negative energy
\item exotic structure of the propagator poles
\end{itemize}
can be reconciled with each other and the known physical principles. We summarize the contents of the 
Sections  \ref{sec:prop}-\ref{sec:field} and Appendix  \ref{sec:app2} by the following interdependent observations:

1) The correct sign of the fermionic lagrangian is not obvious in and of itself. One does not expect trouble in a fermionic theory because of the sign of the lagrangian, because the  Lagrangian 
is first order in derivatives and there are in any case solutions with both positive and negative frequencies. 
One should contrast that with bosonic Lagrangians where the sign of the kinetic term is the same for positive and negative solutions, and therefore the wrong sign such as for a scalar field, $L=-(\partial_\mu \phi)^2$ is certain to cause loss of positivity of both energy and probability.

2) Difficulties with canonical quantization had been caused by the particular definition of conjugate field: for fermions the conjugate field should be always regarded as the hermitian conjugate $\psi^\dagger$ regardless of the lagrangian. This is suggested by the definition of the Berezin fermionic path integral, and more basically by the fact that $ \{  \psi_a (x) , \psi^\dagger_b(y) \} = \Pi_{ab} \, \delta(x-y) $ is positive definite automatically since the spin sum matrix $\Pi_{ab} = \sum u \, u^\dagger + v \, v^\dagger$ always is, moreover in the non-degenerate cases $\Pi_{ab} = \delta_{ab}$. This is to say that the definition $\pi = \tfrac {\delta L}{\delta \dot{\psi} } = -i\,\psi^\dagger \, \gamma_0 \, \Gamma_0$ should be abandoned for the general fermionic systems and in that sense we agree with the results of Sudarshan, Johnson and Munczek who found the non-positive definiteness of the matrix $\gamma_0 \Gamma_0$ to be untenable in conjunction with such definition of $\pi$. With the correct definition, the equal-time anti-commutator is positive definite, and the Lorentz covariant anti-commutator is causal in the sense that it vanishes for space-like separation. The non-definiteness of $\gamma_0 \Gamma_0$ nevertheless makes profound consequences in the propagator, and in the Section \ref{sec:field} we reconcile the expression for the  two-point  time-ordered product obtained by the canonical method with the momentum space propagator obtained in the path-integral formalism.

3) 
The energy is calculated as the expectation value of the energy-momentum tensor $E=\bra{\Omega}\,T_{00}\,\ket{\Omega}$ and the expression is found to differ in sign for the spin 1/2 sector from the sign naively expected, due again to the non-definiteness of $\gamma_0 \Gamma_0$. We argued that the choice of the highest-weight vacuum which is consistent with all the other clues can be made so as to make energy $E > 0$ positive for all physical excitations. 

The choice of the highest-weight vacuum is obviously allowed in the fermionic case, unlike the bosonic case where it 
leads to excitations with negative norm. This is argued in Appendix \ref{sec:app2} to lead to expanded possibilities for the pole structure of the two-point function for fermions precisely of the needed kind, see the mini-discussion of the pole structure below.

4) 
 The pole locations of the multi-particle fermionic systems have not been examined in detail previously.  Specifically, for the two-point function it generally follows that the positivity of the probability $\langle {p} \ket{p}$ requires the residue to be positive at the pole corresponding to the time-forward propagating particle (and negative at the time-backward propagating anti-particle pole). The presented theory satisfies this fundamental requirement. Our fermionic propagator is only unusual in that its positive frequency spin 1/2 pole lies above the real axis. The position of the pole above the axis implies that the corresponding amplitude propagates backward in time. It is then natural that the residue at such pole is found to be negative. 
%
%
%
%
One could also say that the physics comes out ok because the two-point function is still equal to the retarded minus advanced potential, for both spin 3/2 and spin 1/2 particles. The physically unacceptable situation would indeed result if it was found to be of the advanced minus the retarded form. 

\begin{acknowledgments}

The material Appendix \ref{sec:app2} was worked out in collaboration with Edna Cheung.
The author would like to thank 
Edna Cheung and George Savvidy for encouragement and many enlightening discussions. This research was supported in part by NSFC grant No.~10775067 and the Research Links Programme of the Swedish Research Council under contract No.~348-2008-6049..

\end{acknowledgments}

\appendix
\section{Spectral representation for fermionic fields}
\label{sec:app2}
We set out to compute the two-point function in an arbitrary theory with a fundamental or composite fermionic field following Kallen and Lehmann who proved their theorem in the case of a single bosonic field. The main point of interest will be the fact that in the fermionic theory some crucial matrix elements may have contribution either from positive or negative frequency modes of the field without leading to the violation of unitarity. In the end, we obtain a new result, namely that in the fermionic theory 
the propagator $-i/(p^2 -m^2 - i\epsilon)$ is compatible with general principles of QFT in addition to the usual $+i/(p^2 -m^2 + i\epsilon)$. Both are weighted by the positive real spectral function $\rho(m^2)$, which in principle includes contributions
from all orders of perturbation theory but the derivation itself does not depend on an expansion in powers of the coupling. We do not assume any particular Lorentz structure for the field which means that in any specific theory one should afterwards sum over the spin states. 
With this caveat, the two-point function is
\be
G(x,y) = \bra{\Omega} \, T \, \psi(x) \, \psi^\dagger(y) \, \ket{\Omega}~.
\ee
Here, $\ket{\Omega}$ is the true vacuum of the theory about which we shall only make the assumption of Lorentz invariance.

A complete basis of states involves a sum over multiple particle and anti-particle states:
\be
\openone = \ket{\Omega} \, \bra{\Omega} + \sum_{\lambda, s} \int \frac{d^3p}{8\pi^3} ~ \frac 1 {2\, E_p(\lambda)}\,  \ket{\lambda_p,s}\, \bra{\lambda_p,s}
\ee
where $p$ is the total momentum and $s$ is some abstract label enumerating
We now assume $x_0 > y_0$, and insert the complete basis of states into the expression for the two-point function:
\be
G(x,y) = \sum_{\lambda, s}  \int \frac{d^3p}{8\pi^3} ~  \frac 1 {2\, E_p(\lambda)}~
  \bra{\Omega} \, \psi(x) \, \ket{\lambda_p,s}\,
 \bra{\lambda_p,s} \, \psi^\dagger(y)\,  \ket{\Omega}~~.
\ee
First, we consider the standard case, when the matrix element $\bra{\Omega} \, \psi(x)\,  \ket{\lambda_p,s}$ is nonzero due to
 the positive frequency part $\psi_+(x)$,
\be
\bra{\Omega} \, \psi(x) \, \ket{\lambda_p,s} = \e{-ipx} \, \bra{\Omega} \, \psi_+(0) \, \ket{\lambda_p,s}
\ee
with $p_0 > 0 $ and,
\be
\bra{\lambda_p,s} \, \psi^\dagger(y) \, \ket{\Omega} = \e{+ipy} \, \bra{\lambda_p,s} \, \psi_+(0)^\dagger \, \ket{\Omega}
\ee
%
Furthermore,
\be
\bra{\Omega} \, \psi_+(0) \, \ket{\lambda_p,s}^* = \bra{\lambda_p,s} \, \psi_+(0)^\dagger \, \ket{\Omega}~~.
\ee
Then follows the result
\bea
G(x,y) = \sum_{\lambda, s}  \int \frac{d^3p}{8\pi^3} ~ \frac{\exp{(-i\, E_p\, (x_0-y_0) + i\, \vec{p}\, (\vec{x}-\vec{y}) )}  }{2\, E_p}
\vert \bra{\Omega} \psi_+(0) \ket{\lambda_p,s} \vert^2\nn
G(x,y) = \sum_{\lambda, s}  \int \frac{d^4p}{(2\pi)^4} ~ \e{-ip(x-y)} \, \frac{i}{p^2 - m_\lambda^2 + i\, \epsilon} ~
\vert \, \bra{\Omega} \psi_+(0) \ket{\lambda_p,s} \, \vert^2~~.
\eea
When introducing the integration over $p_0$, we must close the contour at $\infty$ from below, and we want to pick up the positive frequency pole $p_0=+E_p$, therefore the sign of $\epsilon$ must be as it is shown.

In the case of $y_0 > x_0$, the contribution is from the other pole and the contour is closed from above, but the final result is identical, meaning that the two-point function is analytical in momenta.

Repeated application of the same procedure, i.e. of inserting a complete basis of states, can be used to extract external legs of the n-point function and obtain the S-matrix, the LSZ theorem.

The new possibilities arise if it so happens that the true vacuum is annihilated by the negative frequency part. This, for example, can happen already in the free theory if the energy has the opposite sign to the usual:
\be
E = - \sum E_p \, ( a^\dagger_p \, a_p + b^\dagger_p \, b_p ) ~.
\ee
In condensed matter it may happen that the lowest energy state is the filled state for some momenta and the empty state for other momenta \cite{deWitt}, giving rise in particular to the Fermi surface; in relativistic theory it is impossible for a crossover to happen for some finite value of momentum between the energy of the filled and empty state. It has been previously overlooked that it is impossible to guarantee that the lowest-weight state is the lowest energy state; the main body of this paper provides a compelling example of a theory where one must choose the highest-weight vacuum for some particle species (spin 1/2) and the usual lowest-weight vacuum for another (spin 3/2). An explanation for why this is generally disallowed in the bosonic case is presented at the end of this Appendix.

We now include the case when the true vacuum is annihilated by the negative frequency part, 
\[
 \bra{\Omega} \, \psi(x) \, \ket{\lambda_p,s} = \bra{\Omega} \, \psi_-(0) \, \ket{\lambda_p,s}\, \e{ipx}~~,
\]
so that for $x_0 > y_0$:
\be
\bra{\Omega} \, \psi(x) \, \psi^\dagger(y) \, \ket{\Omega} = 
 \sum_{\lambda, s}  \int \frac{d^3p}{8\pi^3} ~  \frac 1 {2\, E_p}\, 
 \vert \, \bra{\Omega} \, \psi(0) \, \ket{\lambda_p,s}\, \vert^2~\, \e{ip(x-y)}
\ee
At infinity,  the imaginary part of $p_0$ must be positive $ \Im(p_0) >0$ for the contribution from infinity to vanish, so we must close the contour from above but we still want to pick up the pole so it must lie above the real axis.
%
Finally, with $p_0$ integration inserted,
\be
G(x,y) = \sum_{\lambda}  \int \frac{d^4p}{(2\pi)^4} ~ \e{ip(x-y)} \, \frac{-i}{p^2 - m_\lambda^2 - i\epsilon} ~ \rho(\lambda)\,\Pi_p
\ee
where $\rho(\lambda) \, \Pi_p = \sum_s \vert \bra{\Omega} \, \psi_+(0)\,  \ket{\lambda_p,s} \vert^2 $. Here, $\Pi_p$ is a hermitian matrix with positive eigenvalues whose precise form depends on which representation the field falls into, and is universal in the sense that it can be obtained by Lorentz transformation from some fixed frame; $\rho(\lambda)$ is a positive real spectral density.
An isolated single-particle mass state contributes $\Pi_p\,\delta(\lambda -m)$.

The results can be summarised by stating that the two-point function in an arbitrary theory is a composition of the free propagator with a positive spectral density:
\[
G(x,y) = \int_0^\infty d\mu ~ \rho(\mu) ~ D_F(x-y, \mu)
\]

\[
D_F(x,0) = \int \frac{d^4p}{(2\pi)^4} ~ \, \e{-ipx} \, G(p) 
\]
so that in the standard case
\[
G_F(p) = \frac{iZ}{p^2 - m^2 + i\epsilon}~,
\]
and in the second case
\[
G_{K}(p) = \frac{-iZ}{p^2 - m^2 - i\epsilon}~.
\]
QFT textbooks often mention that there are four possibilities for pole arrangements. Only three are usually discussed: the pure retarded, the pure advanced, and the Feynman propagator. We have arrived at the necessity to allow also the fourth possibility, in conjunction with an overall minus sign, which makes this K-propagator physically acceptable in the fermionic case.

For bosonic fields, 
the alternative vacuum does indeed arise whenever it is necessary to quantize with a negative kinetic energy, for example the $A_0$ component of the photon field. This was first quantized correctly by Gupta and Bleuler by the device of ``negative metric":
\[
a_i \, \ket{0} =0~~,~~~ a_0\, \ket{0} =0
\]
but
\[     [a_i , a^\dagger_j] = \delta_{ij} ~,~~   [a_0 , a^\dagger_0] = {-}1   ~. \]
Another way to say it is that the commutation relation is left standard but the vacuum is chosen to be destroyed by the creation operator. The $A_0$ is then in the highest-weight vacuum. 
The difference from the fermionic case is that the ``excited" states of the highest-weight bosonic vacuum such as $ \ket{{-}1} = a^\dagger_0 \ket{0}$ have negative norm: 
\[ \langle {{-}1}\ket{{-}1}  = \bra{0} \, a_0 \, a^\dagger_0 \, \ket{0} = \bra{0} \, ( - 1 +  a^\dagger_0 \,a_0) \,\ket{0} = -1 ~~. \] 
This means that the highest weight vacuum is generally disallowed in the bosonic case, except in very special circumstances. For the photon field the problem is solved by imposing the subsidiary physical conditions, such that negative norm states do not appear but null-states do; these are then argued to have vanishing scattering amplitudes.  

On the contrary, in the fermionic theory the highest-weight vacuum, has positive norm excitations $ \ket{0} = a \,\ket{1}$ due to 
\[
 \langle {0}\ket{0}  = \bra{1}\, a^\dagger \, a \, \ket{1} = \bra{1}\, ( 1 -  a_0 \,a^\dagger_0)\, \ket{1} = 1 \] 
Obviously, in the fermionic case the choice of the vacuum is at the same time capable of rendering the energy of physical excitation positive, unlike the bosonic case.

 It is worth to mention that in multi-component theories one cannot necessarily choose the sign of $\epsilon$ at will. For example, to preserve Lorentz invariance, the sign of $\epsilon$ cannot be chosen separately for $A_0$ relative to the other components, so that the propagator has the unphysical pole $S=\tfrac{-i\,g_{\mu\nu}}{p^2 +i\,\epsilon}$.  In the specific theory considered in the main body of the paper, the sign of  $\epsilon$ in the spin 1/2 sector also cannot be chosen separately from the spin 3/2 sector, but the additional restriction $M > 0$ results in the structure of the poles which respects unitarity.

\vfill

\end{document}